\documentstyle[prc,aps,twocolumn,amsmath,graphicx]{revtex}

\begin{document}
\draft

\title{In-medium broadening of nucleon resonances}

\author{
J. Lehr\thanks{E-mail: Juergen.Lehr@theo.physik.uni-giessen.de}
and
U. Mosel\thanks{E-mail: mosel@physik.uni-giessen.de}
}

\address{Institut f\"ur Theoretische Physik\\ Universit\"at
Giessen\\ D-35392 Giessen, Germany}

\date{\today}

\maketitle

\begin{abstract}
We analyze the effects of an in-medium broadening of nucleon
resonances on the exclusive photoproduction of mesons on nuclei as
well as on the total photoabsorption cross sections in a transport
calculation. We show that the resonance widths observed in
semi-inclusive photoproduction on nuclei are insensitive to an
in-medium broadening of nucleon resonances. This is due to a
simple effect: the sizeable width of the nuclear
surface and Fermi motion.
\end{abstract}

One of the most obvious in-medium effects is the observed
disappearance of the resonance structures in the second and third
resonance region in the total photoabsorption cross section on
nuclei \cite{Bianchi}. Kondratyuk et al. \cite{Kondratyuk} and
Alberico et al.\ \cite{Alber} have shown that this disappearance
can be explained by an extra in-medium width of about 0.3 GeV for
the higher resonances. Two reasons for this large width have been
discussed. In \cite{Effe,Lehr} we have pointed out that the total
in-medium width of the $D_{13}(1520)$ resonance could reflect a
lowering of the $\rho$ meson spectral strength. A lowering of the
$\rho$ meson strength has been predicted by various hadronic
models and QCD sum rule analyses (for a summary and review see
\cite{Baryon}). Also effects of collisional broadening of this
resonance due to collisions with other baryons have been
considered \cite{Kondratyuk,Alber,Effe}. Both effects lead in
lowest order in the density to an extra in-medium resonance width
that is proportional to the nuclear density $\rho$
\cite{Effe,Lehr,Rapp}.

In a recent paper Krusche et al.\cite{Krusche} discuss their data
on inclusive and semi-inclusive single $\pi^0$ production in the
second resonance region. After subtracting a smooth background
from the observed cross section these authors obtain a resonance
contribution which they attribute to the $D_{13}(1520)$ nucleon
excitation. This resonant part shows a width that is compatible
with the free width of the $D_{13}(1520)$ when smeared over the
Fermi momentum of the nucleons in the target nucleus (see Fig.~5
in \cite{Krusche}). This result seems to present a problem for
explanations of the disappearance of this resonance in the
photoabsorption experiments \cite{Bianchi} in terms of an
in-medium broadening of the $D_{13}$ resonance. Raising this
problem has been the central point of Ref.~\cite{Krusche}.

In this letter we show the first analysis of these data in a
transport theoretical coupled channel BUU (CBUU) calculation. The
results of this calculation exhibit a remarkable, unexpected
insensitivity of the width of the resonant contribution to the
single $\pi^0$ cross section to the strength of the in-medium
broadening. We then present a very simple explanation of this observation. 
The purpose of this
letter is not to provide a perfect reproduction of the data of
\cite{Krusche}, but instead to help solving the apparent dilemma
raised in Ref.~\cite{Krusche}.

We start out by presenting results of a CBUU calculation based on
the method explained in detail in \cite{Effe,Lehr,Effe1}. The
calculations use a spectral function for the nucleon resonance in
the relativistic form
\begin{equation}     \label{spect}
A(s) = \frac{s\Gamma_{\textrm{tot}}(s)}{(s - M^2)^2 + s 
\Gamma_{\rm tot}^2(s)} ~.
\end{equation}
Here $M$ is the resonance mass and $\Gamma_{\rm tot}$ its total
width, which is given as a sum over the total decay width and an
in-medium, density-dependent width $\Gamma_{\textrm{med}}$
\begin{equation}         \label{gammatot}
\Gamma_{\rm tot} = \Gamma_{\rm decay}(s) + \Gamma_{\rm med} ~.
\end{equation}
The in-medium width for the $D_{13}(1520)$ resonance is taken as
\begin{equation}    \label{Gamma}
\Gamma_{\textrm{med}} = \Gamma_{\rm m} \frac{\rho(r)}{\rho_0} ~,
\end{equation}
as is, e.g., appropriate for a width that originates in two body
collisions of the resonance with nucleons in the nuclear medium.
Actually, the precise mechanism that leads to this density
dependent width is irrelevant for the arguments to follow; also
the mechanism described in the first paragraph ($\rho$ mass
lowering) leads in lowest order to a width proportional to the
density $\rho$. The calculations have been done both with and
without this in-medium width. Increasing the width parameter
$\Gamma_{\rm m}$ both broadens the resonance and lowers its
amplitude.

\begin{figure}
\begin{center}
\vspace{-0.5cm}
\includegraphics[width=9cm]{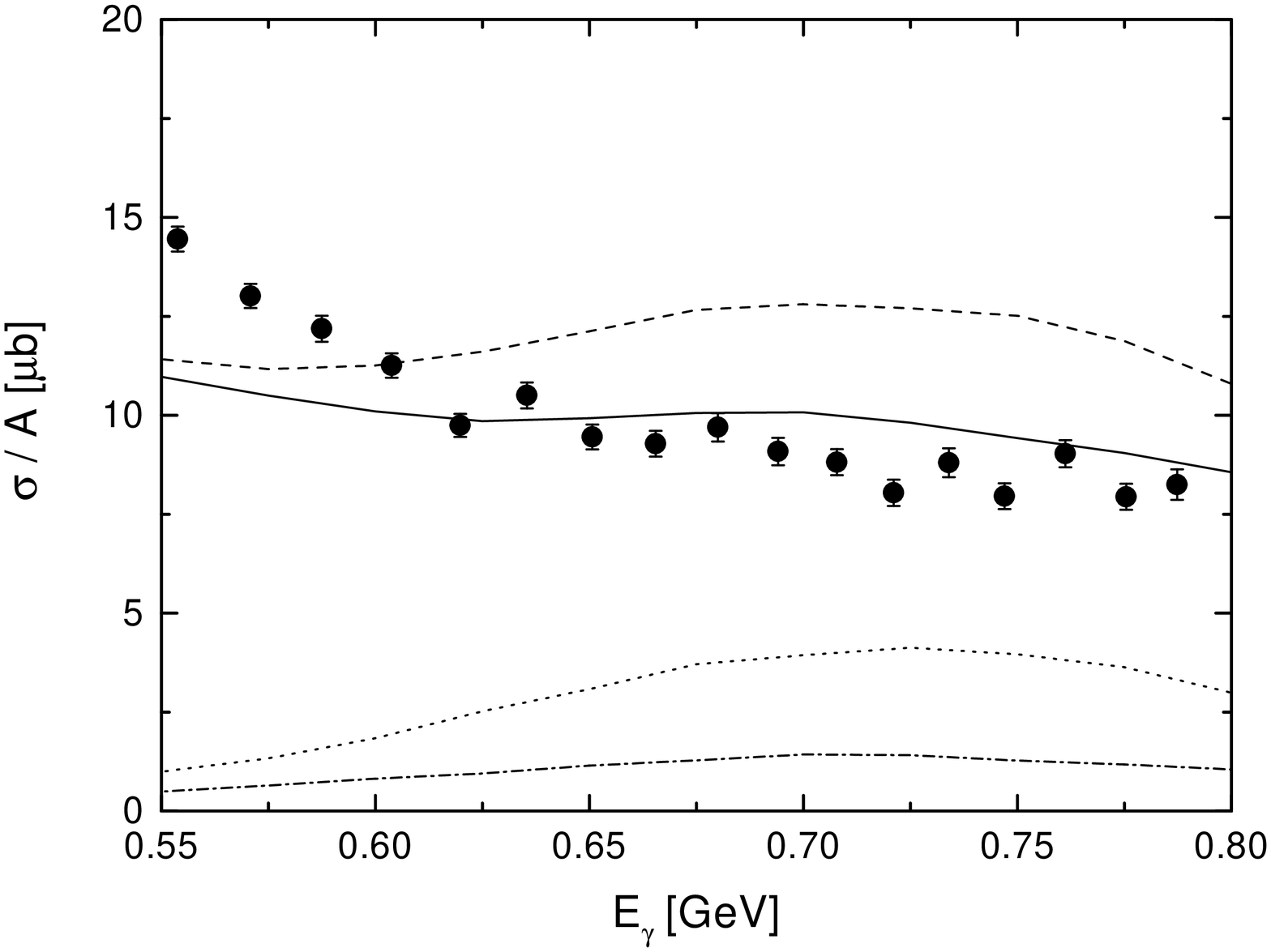}
\end{center}
\caption{Calculated cross section for semi-inclusive single
$\pi^0$ photoproduction on $^{40}\textrm{Ca}$ in comparison to the
data of \protect\cite{Krusche}. The same missing energy cuts as in
\protect\cite{Krusche} have been applied.
Shown are results both without
(dashed) and with an in-medium broadening ($\Gamma_{m} = 0.3$ GeV)
(solid line) and the contribution of the $D_{13}(1520)$ alone for
both widths (dotted and dash-dotted lines).} \label{BUUFig}
\end{figure}

The lowering can clearly be seen in Fig.~\ref{BUUFig}, where we
show the results of this calculation for semi-inclusive single
$\pi^0$ production in comparison to the data of Ref.~\cite{Krusche},
and in Fig.~1 of Ref.~\cite{Krusche} where a
comparison of total inclusive $\pi^0$ photoproduction data with
transport calculations from \cite{Lehr} is shown for various
medium modified widths.

It is seen that the data in the second resonance region are
described reasonably well with an in-medium width parameter
$\Gamma_{\textrm{m}} = 0.3$ GeV \cite{Kondratyuk}; the disagreement at the
lower energies is due to the neglect of two-body absorption with
$\Delta$ degrees of freedom in the calculation \cite{Effe}. The
value $\Gamma_{\rm m} = 0.3$ GeV is hard to justify for a
collisional width alone (as discussed in \cite{Effe,Lehr}); it
serves in this context only as an example for a strong medium
modification and represents the summed effect of a $\rho$ meson
mass shift, collisional broadening and possibly other
density dependent effects.
In the CBUU calculation we can actually follow the history of the
emitted pions and can thus determine which of the observed pions
were emitted last by a $D_{13}(1520)$ resonance. The lower curves
in Fig.~\ref{BUUFig} show this contribution with and without this
in-medium broadening. It can be seen that in the calculation
including the broadening only a very small part ($\sim 10 - 20
\%$) of all single $\pi^0$ in this mass region stems from the
$D_{13}(1520)$ decay; there is a much larger contribution from
other sources (mainly $P_{33}(1232)$, $S_{11}(1535)$ decay and nonresonant
background). This is so although in this energy range the
$D_{13}(1520)$ resonance is dominant for absorption of real
photons on free nucleons, so that the first $(\gamma,\pi)$
reaction proceeds preferably through this resonance. However,
because of the in-medium broadening, this dominance of the
$D_{13}(1520)$ has disappeared in the asymptotic yield.

\begin{figure}
\begin{center}
\vspace{-0.5cm}
\includegraphics[width=9cm]{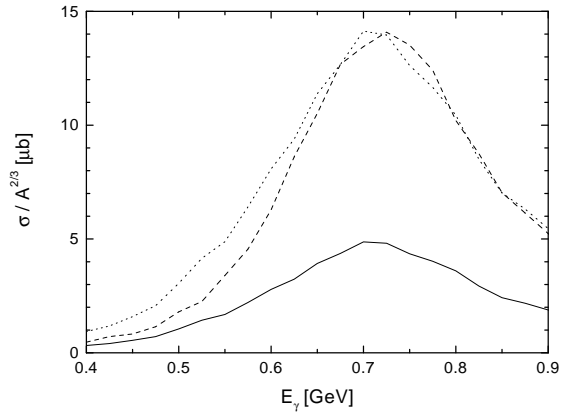}
\end{center}
\caption{Contribution of the $D_{13}(1520)$ to the semi-inclusive
single $\pi^0$ yield in $^{40}{\rm Ca}$ without (dashed line) and
with (solid line) 0.3 GeV in-medium broadening. The dotted line is
identical to the solid line, but scaled up by a factor 2.9 in
order to facilitate the comparison of the widths.} \label{BUUFig2}
\end{figure}

We now analyze the shape of the resonance contribution. Fig.~\ref{BUUFig2}
shows the single $\pi^0$ cross section from the
$D_{13}(1520)$ alone (lower curves in Fig.~\ref{BUUFig}) in some
more detail. As already seen, the cross section drops when the
in-medium width is increased to 0.3 GeV. However, the shape of the
single $\pi^0$ cross section as a function of the photon energy
does not change significantly when the collisional width is
increased. This -- at first sight -- surprising result implies
that in this calculation the total width of the single $\pi^0$
production excitation function is insensitive to collisional
broadening.
A first guess to explain this behavior is obviously ascribing it
to final state interactions. The pions experience strong
reabsorption so that the observed pions can be expected to
originate somewhere in the surface region of the nucleus. The
observed $A^{2/3}$ dependence of the single pion cross section
\cite{Krusche} seems to substantiate this explanation.

We now show, however, that even without reabsorption the width of
the observed resonance distribution is insensitive to an in-medium
broadening of nucleon resonances whereas, on the contrary, total
absorption cross sections directly reflect this broadening.

We write the cross section for the primary production of single
pions from nuclei as a function of incoming photon energy
$E_\gamma$ by simply folding the elementary ($\gamma,\pi$) cross
section on the nucleon with the nuclear phase space distribution
in the Thomas-Fermi approximation. This gives \cite{Effe1}
\begin{eqnarray}              \label{sigma}
\sigma_\pi(E_\gamma) &\sim & \intop_{\rm NV} dr~ r^2
\intop_0^{p_{\rm F}(r)} d^3p~
 \Gamma_\gamma(s(E_\gamma,\vec p))\\ \nonumber
 & &\times ~ A(s(E_\gamma,\vec p)) 
\frac{\Gamma_\pi(s(E_\gamma,\vec p))}{\Gamma_{\textrm{tot}}
(s(E_\gamma,\vec p))}.
\end{eqnarray}
Here $\Gamma_\gamma$ is the in-width of the incoming photon
\cite{Walker}, $\Gamma_\pi$ the one pion decay width, and $s$ is
the square of the invariant resonance mass. The $s$ dependences of
$\Gamma_\pi$ and $\Gamma_{\rm decay}$ are taken from the analysis
of Manley et al. \cite{Manley}. The momentum space integral runs
over the ground state nucleon momenta up to the local Thomas-Fermi
momentum $p_{\rm F}(r)$ and the $r$ space integral extends over
the nuclear volume. The cross section (\ref{sigma}) does not
contain any final state interactions of the produced pions, which
are automatically contained in the BUU results shown in Fig.~\ref{BUUFig}.

We now analyze the resonance shape of the cross section
(\ref{sigma}) as a function of the in-medium width parameter
$\Gamma_{\rm m}$. We perform the calculations for the nucleus
$^{40}\textrm{Ca}$ with a density distribution of the Woods-Saxon
type with parameters given in \cite{Lehr}.

\begin{figure}
\begin{center}
\vspace{-0.5cm}
\includegraphics[width=9cm]{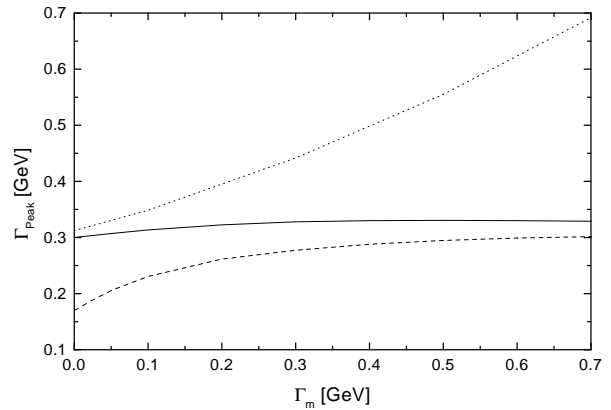}
\end{center}
\caption{Full width at half maximum of the photoproduction cross
section for single pions (\ref{sigma}) for $^{40}{\rm Ca}$ as a
function of the in-medium width $\Gamma_{\rm m}$. The solid line
gives the calculation with Fermi motion included, the dashed line
that without Fermi motion. Also shown is the total FWHM for the
photoabsorption cross section (dotted).} \label{width}
\end{figure}

In Fig.~\ref{width} we show the observed full width $\Gamma_{\rm
peak}$ of the resonance in the one pion photoproduction cross
section $\sigma_\pi(E_\gamma)$ as a function of the in-medium
width $\Gamma_{\rm m}$. The figure contains results of a
calculation both with Fermi motion included and without. It is
seen that the resonance width as a function of the in-medium width
first increases in both cases, as expected. However, in both cases
the widths actually level off as a function of $\Gamma_{\rm m}$;
the realistic case with Fermi motion of the nucleons included
rises by about 10\% from 0.3 GeV for $\Gamma_{\rm m} = 0$ up to
about 0.33 GeV at $\Gamma_{\rm m} = 300$ MeV and then stays
roughly constant. This change is so small that it cannot be
observed in the data.

This behavior is a direct consequence of the form of the cross
section (\ref{sigma}) and, more specifically, the density
dependence of the in-medium width $\Gamma_{\rm med}$
(\ref{Gamma}). The $\rho$ dependence of the in-medium width in
Eqs.~(\ref{Gamma}) and (\ref{sigma}) leads, through its lowering
of the amplitude, to a decrease of contributions from the nuclear
interior. This decrease is stronger closer to the center of the
nucleus and weaker farther out. Thus, increasing the in-medium
width parameter $\Gamma_{\rm m}$ in (\ref{Gamma}) effectively
moves the region of highest sensitity out from the maximum of
$\rho(r) r^2$ to even larger radii with smaller density, so that
the observed in-medium width stays approximately constant. In a
(fictitious) nucleus with constant density the total width does
not level off, but increases with $\Gamma_{\rm m}$; the same holds
if the in-medium width does not depend on density.

We note that the detailed form of the $s$ dependence of the total
decay and pion widths $\Gamma_{\rm decay}$ and $\Gamma_\pi$ is not
essential for this result. Taking constant values for both widths
also leads to a levelling off, although at a somewhat higher
value, of the total width. Thus, purely geometrical effects alone
lead to the observed near independence of the resonance width on
the in-medium width; the reabsorption neglected in (\ref{sigma})
can only enhance this behavior since it also acts to shift the
'sensitivity region' further out.

Semi-inclusive meson photoproduction experiments are thus nearly
insensitive to an in-medium broadening of the nucleon resonances
as far as their shape is concerned; only their amplitude is
diminished. On the contrary, the width of the resonant part of the
total photoabsorption cross section is quite sensitive to an
in-medium broadening. In this case, Eq.~(\ref{sigma}) has to be
modified such that the factor $\Gamma_\pi$ in the integrand in
(\ref{sigma}) is replaced by $\Gamma_{\rm tot}$, which includes as
one of its terms the width $\Gamma_{\rm med}$ (see (\ref{Gamma},
(\ref{gammatot}))). For large $\Gamma_{\textrm{med}}$ this latter
term becomes dominant and counteracts the corresponding term in
the denominator of the Breit-Wigner distribution in (\ref{sigma}).
Indeed, when plotting the observed width of the absorption cross
section  vs.\ $\Gamma_{\rm m}$ one finds a continuous increase
without any levelling off (see Fig.~\ref{width}, dotted line).

The broadening of nucleon resonances can thus be seen in the
photoabsorption experiments, but not in the more exclusive pion
production experiments. Also, a somewhat larger effect of
in-medium broadening remains in the invariant mass distribution of
the resonance, constructed from its decay products, because here
the Fermi motion effect is absent (see dashed curve in Fig.~3)
and the width in the $\sqrt{s}$ distribution can directly be seen.
Of course, all different reaction channels on the nucleon are
determined by one and the same total width of the resonance.
However, in experiments with nuclei the different reactions sample
the resonance at different densities and thus see different total
widths. This different sensitivity of the semi-inclusive and the
fully inclusive processes is a simple consequence of nuclear
geometry. This finding is in line with the empirically observed
$A$ dependence of the cross sections. Whereas photoabsorption
cross sections scale with $A$ for nuclei heavier than Carbon
\cite{Bianchi}, the semi-inclusive and inclusive $\pi^0$ cross
sections scale with $A^{2/3}$ \cite{Krusche}, thus indicating a
volume and a surface effect, respectively.

Finally, we mention that the width of the density averaged
spectral function $A(s)$ increases with $\Gamma_{\textrm{m}}$, but
considerably more slowly than that of the absorption cross section.
In addition, it is free of any Fermi motion effects. Experiments
that reconstruct the invariant mass distribution are, therefore,
sensitive to in-medium broadening, if the final state interactions can
be neglected.

In summary, the width of the resonant part of the semi-inclusive
single $\pi^0$ photoproduction in the second resonance region is
found to be insensitive to an in-medium broadening of the
resonance, in contrast to photoabsorption. This explains the
results of Ref.~\cite{Krusche} and solves the apparent problem
discussed there. It also has an important implication for planned
photoproduction experiments on nuclei that are motivated by the
search for in-medium changes of hadronic properties.

\acknowledgments The authors would like to thank T. Falter for
many valuable discussions and comments. Furthermore, they would
like to thank V. Metag for stimulating discussions and B. Krusche
for many critical, helpful remarks on this manuscript. This work
was supported by DFG.



\begin{thebibliography}{99}

\bibitem{Bianchi} N. Bianchi et al., Phys. Rev. {\bf C54}, 1688 (1996).
\bibitem{Kondratyuk} L. A. Kondratyuk et al., Nucl.Phys. {\bf A579}, 453
(1994).
\bibitem{Alber} W.M. Alberico, G. Gervino and A. Lavagno, Phys. Lett.
{\bf B321}, 177 (1994)
\bibitem{Effe} M. Effenberger et al., Nucl. Phys. {\bf A613}, 353 (1997).
\bibitem{Lehr} J. Lehr, M. Effenberger and U. Mosel, Nucl. Phys.
{\bf A671}, 503 (2000).
\bibitem{Baryon} U. Mosel, in {\it Baryons'98, Proc. 8th Int. Conf. Structure of
Baryons}, World Scientific, Singapore, 1999, p. 629.
\bibitem{Rapp} R. Rapp et al., Phys. Lett. {\bf B417}, 1 (1998)
\bibitem{Krusche} B. Krusche et al., Phys. Rev. Lett. {\bf 86} 4764
(2001).
\bibitem{Manley} D.M. Manley and E.M. Saleski, Phys. Rev.
{\bf D45}, 4002 (1992).
\bibitem{Effe1} M. Effenberger et al.,
Nucl. Phys. {\bf A614}, 501 (1997).
\bibitem{Walker} R. L. Walker, Phys. Rev. {\bf 182}, 1729 (1969).

\end{thebibliography}
\end{document}